\input epsf
%\input midimac
%%%%%%%%%%%%%%%%%%%  tex macros for preprints, cm version %%%%%%%%%%%%%%
%                     (P. Ginsparg, last updated 9/91)
%                if confused, type `b' in response to query
%
%---------------------------------------------------------------------%
%% site dependent options:
%% \unredoffs and \redoffs define horizontal and vertical offsets
%% respectively for unreduced and reduced modes. \speclscape defines
%% the \special{} call that sets printer to landscape (sideways) mode.
%% from standard set below, leave uncommented as appropriate or redefine
%
%%% next 400dpi
%\def\unredoffs{} \def\redoffs{\voffset=-.31truein\hoffset=-.48truein}
%\def\speclscape{\special{landscape}}
%
%%% apple lw
\def\unredoffs{} 

%
%%% qms lasergrafix:
%\def\unredoffs{} \def\redoffs{\voffset=-.4truein\hoffset=.125truein}
%\def\speclscape{\special{qms: landscape}}
%
%%% saclay A4 paper:
%\def\unredoffs{\hoffset-.14truein\voffset-.2truein}
%\def\redoffs{\voffset=-.45truein\hoffset=-.21truein}
%\def\speclscape{\special{landscape}}
%
%---------------------------------------------------------------------%
%
\newbox\leftpage \newdimen\fullhsize \newdimen\hstitle \newdimen\hsbody
\tolerance=1000\hfuzz=2pt
\catcode`\@=11 % This allows us to modify PLAIN macros.
%\def\bigans{b }
%\message{ big or little (b/l)? }\read-1 to\answ
%
%\ifx\answ\bigans\message{(This will come out unreduced.}
\magnification=1200\unredoffs\baselineskip=16pt plus 2pt minus 1pt
\hsbody=\hsize \hstitle=\hsize %take default values for unreduced format
%
%\else\message{(This will be reduced.} \let\l@r=L
%\magnification=1000\baselineskip=16pt plus 2pt minus 1pt \vsize=7truein
%\redoffs \hstitle=8truein\hsbody=4.75truein\fullhsize=10truein\hsize=\hsbody
%
%\output={\ifnum\pageno=0 %%% This is the HUTP version
%  \shipout\vbox{\speclscape{\hsize\fullhsize\makeheadline}
%    \hbox to \fullhsize{\hfill\pagebody\hfill}}\advancepageno
%  \else
%  \almostshipout{\leftline{\vbox{\pagebody\makefootline}}}\advancepageno
%  \fi}
%\def\almostshipout#1{\if L\l@r \count1=1 \message{[\the\count0.\the\count1]}
%      \global\setbox\leftpage=#1 \global\let\l@r=R
% \else \count1=2
%  \shipout\vbox{\speclscape{\hsize\fullhsize\makeheadline}
%      \hbox to\fullhsize{\box\leftpage\hfil#1}}  \global\let\l@r=L\fi}
%\fi
%---------------------------------------------------------------------
%
\newcount\yearltd\yearltd=\year\advance\yearltd by -1900

%
%      restores pagenumbers
%
%       use following instead of \Date on the preliminary draft,
%       puts date/time on each page in big mode, writes labels in margins

\def\draftmode{\message{ DRAFTMODE }\def\draftdate{{\rm preliminary draft:
\number\month/\number\day/\number\yearltd\ \ \hourmin}}%
\headline={\hfil\draftdate}\writelabels\baselineskip=20pt plus 2pt minus 2pt
 {\count255=\time\divide\count255 by 60 \xdef\hourmin{\number\count255}
  \multiply\count255 by-60\advance\count255 by\time
  \xdef\hourmin{\hourmin:\ifnum\count255<10 0\fi\the\count255}}}
%       use \nolabels to get rid of eqn, ref], and fig labels in draft mode
\def\nolabels{\def\wrlabeL##1{}\def\eqlabeL##1{}\def\reflabeL##1{}}
\def\writelabels{\def\wrlabeL##1{\leavevmode\vadjust{\rlap{\smash%
{\line{{\escapechar=` \hfill\rlap{\sevenrm\hskip.03in\string##1}}}}}}}%
\def\eqlabeL##1{{\escapechar-1\rlap{\sevenrm\hskip.05in\string##1}}}%
\def\reflabeL##1{\noexpand\llap{\noexpand\sevenrm\string\string\string##1}}}
\nolabels
%
% tagged sec numbers
\global\newcount\secno \global\secno=0
\global\newcount\meqno \global\meqno=1
\def\newsec#1{\global\advance\secno by1\message{(\the\secno. #1)}
%\ifx\answ\bigans \vfill\eject \else \bigbreak\bigskip \fi  %if desired
\global\subsecno=0\eqnres@t\noindent{\bf\the\secno. #1}
\writetoca{{\secsym} {#1}}\par\nobreak\medskip\nobreak}
\def\eqnres@t{\xdef\secsym{\the\secno.}\global\meqno=1\bigbreak\bigskip}
\def\sequentialequations{\def\eqnres@t{\bigbreak}}\xdef\secsym{}
\global\newcount\subsecno \global\subsecno=0
\def\subsec#1{\global\advance\subsecno by1\message{(\secsym\the\subsecno. #1)}
\ifnum\lastpenalty>9000\else\bigbreak\fi
\noindent{\it\secsym\the\subsecno. #1}\writetoca{\string\quad
{\secsym\the\subsecno.} {#1}}\par\nobreak\medskip\nobreak}
\def\appendix#1#2{\global\meqno=1\global\subsecno=0\xdef\secsym{\hbox{#1.}}
\bigbreak\bigskip\noindent{\bf Appendix #1. #2}\message{(#1. #2)}
\writetoca{Appendix {#1.} {#2}}\par\nobreak\medskip\nobreak}
%
%       \eqn\label{a+b=c}       gives displayed equation, numbered
%                               consecutively within sections.
%     \eqnn and \eqna define labels in advance (of eqalign?)
%
\def\eqnn#1{\xdef #1{(\secsym\the\meqno)}\writedef{#1\leftbracket#1}%
\global\advance\meqno by1\wrlabeL#1}
\def\eqna#1{\xdef #1##1{\hbox{$(\secsym\the\meqno##1)$}}
\writedef{#1\numbersign1\leftbracket#1{\numbersign1}}%
\global\advance\meqno by1\wrlabeL{#1$\{\}$}}
\def\eqn#1#2{\xdef #1{(\secsym\the\meqno)}\writedef{#1\leftbracket#1}%
\global\advance\meqno by1$$#2\eqno#1\eqlabeL#1$$}
%
%                        footnotes
\newskip\footskip\footskip14pt plus 1pt minus 1pt %sets footnote baselineskip
\def\footnotefont{\ninepoint}\def\f@t#1{\footnotefont #1\@foot}
\def\f@@t{\baselineskip\footskip\bgroup\footnotefont\aftergroup\@foot\let\next}
\setbox\strutbox=\hbox{\vrule height9.5pt depth4.5pt width0pt}
\global\newcount\ftno \global\ftno=0
\def\foot{\global\advance\ftno by1\footnote{$^{\the\ftno}$}}
%
%say \footend to put footnotes at end
%will cause problems if \ref used inside \foot, instead use \nref before
\newwrite\ftfile
\def\footend{\def\foot{\global\advance\ftno by1\chardef\wfile=\ftfile
$^{\the\ftno}$\ifnum\ftno=1\immediate\openout\ftfile=foots.tmp\fi%
\immediate\write\ftfile{\noexpand\smallskip%
\noexpand\item{f\the\ftno:\ }\pctsign}\findarg}%
\def\footatend{\vfill\eject\immediate\closeout\ftfile{\parindent=20pt
\centerline{\bf Footnotes}\nobreak\bigskip\input foots.tmp }}}
\def\footatend{}
%
%     \ref\label{text}
% generates a number, assigns it to \label, generates an entry.
% To list the refs on a separate page,  \listrefs
%
\global\newcount\refno \global\refno=1
\newwrite\rfile
\def\ref{[\the\refno]\nref}
\def\nref#1{\xdef#1{[\the\refno]}\writedef{#1\leftbracket#1}%
\ifnum\refno=1\immediate\openout\rfile=refs.tmp\fi
\global\advance\refno by1\chardef\wfile=\rfile\immediate
\write\rfile{\noexpand\item{#1\ }\reflabeL{#1\hskip.31in}\pctsign}\findarg}
%       horrible hack to sidestep tex \write limitation
\def\findarg#1#{\begingroup\obeylines\newlinechar=`\^^M\pass@rg}
{\obeylines\gdef\pass@rg#1{\writ@line\relax #1^^M\hbox{}^^M}%
\gdef\writ@line#1^^M{\expandafter\toks0\expandafter{\striprel@x #1}%
\edef\next{\the\toks0}\ifx\next\em@rk\let\next=\endgroup\else\ifx\next\empty%
\else\immediate\write\wfile{\the\toks0}\fi\let\next=\writ@line\fi\next\relax}}
\def\striprel@x#1{} \def\em@rk{\hbox{}}
\def\lref{\begingroup\obeylines\lr@f}
\def\lr@f#1#2{\gdef#1{\ref#1{#2}}\endgroup\unskip}
\def\semi{;\hfil\break}
\def\addref#1{\immediate\write\rfile{\noexpand\item{}#1}} %now unnecessary
\def\startrefs#1{\immediate\openout\rfile=refs.tmp\refno=#1}
\def\xref{\expandafter\xr@f}\def\xr@f[#1]{#1}
\def\refs#1{\count255=1[\r@fs #1{\hbox{}}]}
\def\r@fs#1{\ifx\und@fined#1\message{reflabel \string#1 is undefined.}%
\nref#1{need to supply reference \string#1.}\fi%
\vphantom{\hphantom{#1}}\edef\next{#1}\ifx\next\em@rk\def\next{}%
\else\ifx\next#1\ifodd\count255\relax\xref#1\count255=0\fi%
\else#1\count255=1\fi\let\next=\r@fs\fi\next}
%

%
% this is ugly, but moore insists
\newwrite\ffile\global\newcount\figno \global\figno=1
\def\fig{fig.~\the\figno\nfig}
\def\nfig#1{\xdef#1{fig.~\the\figno}%
\writedef{#1\leftbracket fig.\noexpand~\the\figno}%
\ifnum\figno=1\immediate\openout\ffile=figs.tmp\fi\chardef\wfile=\ffile%
\immediate\write\ffile{\noexpand\medskip\noexpand\item{Fig.\ \the\figno. }
\reflabeL{#1\hskip.55in}\pctsign}\global\advance\figno by1\findarg}
\def\xfig{\expandafter\xf@g}\def\xf@g fig.\penalty\@M\ {}
\def\figs#1{figs.~\f@gs #1{\hbox{}}}
\def\f@gs#1{\edef\next{#1}\ifx\next\em@rk\def\next{}\else
\ifx\next#1\xfig #1\else#1\fi\let\next=\f@gs\fi\next}
\newwrite\lfile
{\escapechar-1\xdef\pctsign{\string\%}\xdef\leftbracket{\string\{}
\xdef\rightbracket{\string\}}\xdef\numbersign{\string\#}}

\def\writestop{\def\writestoppt{\immediate\write\lfile{\string\pageno%
\the\pageno\string\startrefs\leftbracket\the\refno\rightbracket%
\string\def\string\secsym\leftbracket\secsym\rightbracket%
\string\secno\the\secno\string\meqno\the\meqno}\immediate\closeout\lfile}}
\def\writestoppt{}\def\writedef#1{}
\def\seclab#1{\xdef #1{\the\secno}\writedef{#1\leftbracket#1}\wrlabeL{#1=#1}}
\def\subseclab#1{\xdef #1{\secsym\the\subsecno}%
\writedef{#1\leftbracket#1}\wrlabeL{#1=#1}}
\newwrite\tfile \def\writetoca#1{}
\def\leaderfill{\leaders\hbox to 1em{\hss.\hss}\hfill}
%       use this to write file with table of contents
\def\writetoc{\immediate\openout\tfile=toc.tmp
   \def\writetoca##1{{\edef\next{\write\tfile{\noindent ##1
   \string\leaderfill {\noexpand\number\pageno} \par}}\next}}}
%       and this lists table of contents on second pass

%
\catcode`\@=12 % at signs are no longer letters
%
%       Unpleasantness in calling in abstract and title fonts
\edef\tfontsize{\ifx\answ\bigans scaled\magstep3\else scaled\magstep4\fi}
\font\titlerm=cmr10 \tfontsize \font\titlerms=cmr7 \tfontsize
\font\titlermss=cmr5 \tfontsize \font\titlei=cmmi10 \tfontsize
\font\titleis=cmmi7 \tfontsize \font\titleiss=cmmi5 \tfontsize
\font\titlesy=cmsy10 \tfontsize \font\titlesys=cmsy7 \tfontsize
\font\titlesyss=cmsy5 \tfontsize \font\titleit=cmti10 \tfontsize
\skewchar\titlei='177 \skewchar\titleis='177 \skewchar\titleiss='177
\skewchar\titlesy='60 \skewchar\titlesys='60 \skewchar\titlesyss='60
\def\titlefont{\def\rm{\fam0\titlerm}% switch to title font
\textfont0=\titlerm \scriptfont0=\titlerms \scriptscriptfont0=\titlermss
\textfont1=\titlei \scriptfont1=\titleis \scriptscriptfont1=\titleiss
\textfont2=\titlesy \scriptfont2=\titlesys \scriptscriptfont2=\titlesyss
\textfont\itfam=\titleit \def\it{\fam\itfam\titleit}\rm}
 \ifx\answ\bigans\else scaled\magstep1\fi
\ifx\answ\bigans\else

 \font\absi=cmmi10 scaled\magstep1
\font\absis=cmmi7 scaled\magstep1 \font\absiss=cmmi5 scaled\magstep1
\font\abssy=cmsy10 scaled\magstep1 \font\abssys=cmsy7 scaled\magstep1
\font\abssyss=cmsy5 scaled\magstep1 
\skewchar\absi='177 \skewchar\absis='177 \skewchar\absiss='177
\skewchar\abssy='60 \skewchar\abssys='60 \skewchar\abssyss='60
\fi
\font\ninerm=cmr9 \font\sixrm=cmr6 \font\ninei=cmmi9 \font\sixi=cmmi6
\font\ninesy=cmsy9 \font\sixsy=cmsy6 \font\ninebf=cmbx9
\font\nineit=cmti9 \font\ninesl=cmsl9 \skewchar\ninei='177
\skewchar\sixi='177 \skewchar\ninesy='60 \skewchar\sixsy='60
\def\ninepoint{\def\rm{\fam0\ninerm}% switch to footnote font
\textfont0=\ninerm \scriptfont0=\sixrm \scriptscriptfont0=\fiverm
\textfont1=\ninei \scriptfont1=\sixi \scriptscriptfont1=\fivei
\textfont2=\ninesy \scriptfont2=\sixsy \scriptscriptfont2=\fivesy
\textfont\itfam=\ninei \def\it{\fam\itfam\nineit}\def\sl{\fam\slfam\ninesl}%
\textfont\bffam=\ninebf \def\bf{\fam\bffam\ninebf}\rm}
%
%---------------------------------------------------------------------
%
\def\noblackbox{\overfullrule=0pt}
\hyphenation{anom-aly anom-alies coun-ter-term coun-ter-terms}
\def\inv{^{\raise.15ex\hbox{${\scriptscriptstyle -}$}\kern-.05em 1}}

\def\Dsl{\,\raise.15ex\hbox{/}\mkern-13.5mu D} %this one can be subscripted
\def\dsl{\raise.15ex\hbox{/}\kern-.57em\partial}

 %pound sterling
\def\lspace{\ifx\answ\bigans{}\else\qquad\fi}
\def\lbspace{\ifx\answ\bigans{}\else\hskip-.2in\fi} % $$\lbspace...$$
\def\boxeqn#1{\vcenter{\vbox{\hrule\hbox{\vrule\kern3pt\vbox{\kern3pt
        \hbox{${\displaystyle #1}$}\kern3pt}\kern3pt\vrule}\hrule}}}
\def\mbox#1#2{\vcenter{\hrule \hbox{\vrule height#2in
                \kern#1in \vrule} \hrule}}  %e.g. \mbox{.1}{.1}
%       matters of taste
%\def\tilde{\widetilde} \def\bar{\overline} \def\hat{\widehat}
%
% some sample definitions
  %     curly letters
   
 \def\CH{{\cal H}}

\def\darr#1{\raise1.5ex\hbox{$\leftrightarrow$}\mkern-16.5mu #1}
 %pound sterling

\def\half{{\textstyle{1\over2}}} %puts a small half in a displayed eqn
\def\roughly#1{\raise.3ex\hbox{$#1$\kern-.75em\lower1ex\hbox{$\sim$}}}
\hyphenation{Mar-ti-nel-li}
\def\hp{{\hat p}}

\def\O{{\cal O}}
\def\dm{{\partial\over \partial M}}
\def\gtoy{\gamma_{\rm toy}}
\def\ctoy{\chi_{\rm toy}}

\def\1{\;1\!\!\!\! 1\;}

\def\ie{{\it i.e.}}

\def\etal{{\it et al.}}
\def\rhs{right hand side}

\def\frac#1#2{{{#1}\over {#2}}}
\def\half{\hbox{${1\over 2}$}}
\def\quarter{\hbox{${1\over 4}$}}
\def\smallfrac#1#2{\hbox{${{#1}\over {#2}}$}}

\catcode`@=11 %This allows us to modify plain macros
\def\slash#1{\mathord{\mathpalette\c@ncel#1}}
 \def\c@ncel#1#2{\ooalign{$\hfil#1\mkern1mu/\hfil$\crcr$#1#2$}}
\def\lsim{\mathrel{\mathpalette\@versim<}}
\def\gsim{\mathrel{\mathpalette\@versim>}}
 \def\@versim#1#2{\lower0.2ex\vbox{\baselineskip\z@skip\lineskip\z@skip
       \lineskiplimit\z@\ialign{$\m@th#1\hfil##$\crcr#2\crcr\sim\crcr}}}
\catcode`@=12 %at signs are no longer letters

\def\PR{{\it Phys.~Rev.~}}

\def\NP{{\it Nucl.~Phys.~}}

\def\PL{{\it Phys.~Lett.~}}

\def\SJNP{{\it Sov.~Jour.~Nucl.~Phys.~}}
\def\SPJETP{{\it Sov.~Phys.~J.E.T.P.~}}

\def\JHEP{{\it Jour.~High~Energy~Phys.~}}
\def\vol#1{{\bf #1}}\def\vyp#1#2#3{\vol{#1} (#2) #3}

\def\as{\alpha_s}
\def\ahat{\hat\as}
\def\ahati{{\hat\as^{-1}}}
\def\Mi{{M^{-1}}}
\def\Ni{{N^{-1}}}

\def\ash{\widehat\alpha_s}

%\draftmode
\noblackbox
\pageno=0\nopagenumbers\tolerance=10000\hfuzz=5pt
\baselineskip 12pt
\line{\hfill Edinburgh 2005/23}
\line{\hfill IFUM-868-FT}
\vskip 12pt
\centerline{\titlefont All order Running Coupling}
\vskip 4pt
\centerline{\titlefont BFKL Evolution from GLAP}
\vskip 4pt
\centerline{\titlefont (and vice-versa)}
\vskip 36pt\centerline{Richard D.~Ball$^{(a)}$ and Stefano Forte$^{(b)}$}
\vskip 12pt
\centerline{\it ${}^{(a)}$School of Physics, University of Edinburgh}
\centerline{\it  Edinburgh EH9 3JZ, Scotland}
\vskip 6pt
\centerline {\it ${}^{(b)}$Dipartimento di  Fisica, Universit\`a di
Milano and}
\centerline{\it INFN, Sezione di Milano, Via Celoria 16, I-20133 Milan, Italy}
\vskip 40pt
\centerline{\bf Abstract}
{\narrower\baselineskip 10pt
\medskip\noindent 
We present a systematic formalism for the derivation of the kernel of the BFKL
equation from that of the GLAP equation and
conversely to any given order,  with full
inclusion of the  running of the coupling. 
The running coupling is treated as an operator, reducing the
inclusion  of running coupling effects and their factorization to a
purely algebraic problem. 
We show how the GLAP anomalous dimensions which
resum large logs of $\frac{1}{x}$ can be derived from the
running-coupling BFKL kernel
order by order, thereby obtaining a constructive all-order proof of
small $x$ factorization. We check this result by explicitly
calculating the running coupling contributions to GLAP anomalous
dimensions up to next-to-next-to leading order.
We finally derive an explicit expression for  
BFKL kernels which resum large logs of $Q^2$ up to next-to-leading
order from the corresponding GLAP kernels,
thus making possible a consistent collinear improvement of the BFKL
equation up to the same order. 
}
\vfill
\line{December 2005\hfill}
\eject \footline={\hss\tenrm\folio\hss}
%%%%%%%%%%%%%%%%%%%%%%%%%%%%%%%%%%%%%%%%%%%%%%%%%%%%%%%%%%%%%%%%%
% references
%%%%%%%%%%%%%%%%%%%

\lref\bch{E.~Eriksen, {\it J. Math. Phys.}\vyp{9}{1969}{790}}
\lref\brus  {R.~D.~Ball and S.~Forte,
  %``Corrections at small x,''
  {\tt hep-ph/9805315.}
  %%CITATION = HEP-PH 9805315;%%
}
\lref\fal{P.~Falgari, {\it Laurea Thesis}, Milan University, April 2005}
\lref\mar{S.~Marzani, {\it Laurea Thesis}, Milan University, April 2005}
\lref\marfal{R.~D.~Ball, P.~Falgari, S.~Forte and S.~Marzani, {\it in
  preparation} }
\lref\glap{
V.N.~Gribov and L.N.~Lipatov,
\SJNP\vyp{15}{1972}{438}\semi  %%CITATION = YAFIA,15,781;%%
L.N.~Lipatov, \SJNP\vyp{20}{1975}{95}\semi    %%CITATION = YAFIA,20,95;%%
G.~Altarelli and G.~Parisi,
\NP\vyp{B126}{1977}{298}\semi  %%CITATION = NUPHA,B126,298;%%
see also
Y.L.~Dokshitzer,
{\it Sov.~Phys.~JETP~}\vyp{46}{1977}{691}.} %%CITATION = SPHJA,46,691;%%
\lref\nlo{G.~Curci, W.~Furma\'nski and R.~Petronzio,
\NP\vyp{B175}{1980}{27}\semi %%CITATION = NUPHA,B175,27;%%
E.G.~Floratos, C.~Kounnas and R.~Lacaze,
\NP\vyp{B192}{1981}{417}.} %%CITATION = NUPHA,B192,417;%%
\lref\nnlo{S.A.~Larin, T.~van~Ritbergen, J.A.M.~Vermaseren,
\NP\vyp{B427}{1994}{41}\semi  %%CITATION = NUPHA,B427,41;%%
S.A.~Larin \etal, \NP\vyp{B492}{1997}{338}.} %%CITATION = HEP-PH 9605317;%%
\lref\bfkl{L.N.~Lipatov,
\SJNP\vyp{23}{1976}{338}\semi %%CITATION = SJNCA,23,338;%%
 V.S.~Fadin, E.A.~Kuraev and L.N.~Lipatov,
\PL\vyp{60B}{1975}{50}; %%CITATION = PHLTA,B60,50;%%
 {\it Sov. Phys. JETP~}\vyp{44}{1976}{443}; %%CITATION = SPHJA,44,443;%%
\vyp{45}{1977}{199}\semi %%CITATION = SPHJA,45,199;%%
 Y.Y.~Balitski and L.N.Lipatov,
\SJNP\vyp{28}{1978}{822}.} %%CITATION = SJNCA,28,822;%%
\lref\CH{
S.~Catani and F.~Hautmann,
\PL\vyp{B315}{1993}{157}; %%CITATION = PHLTA,B315,157;%%
\NP\vyp{B427}{1994}{475}.} %%CITATION = NUPHA,B427,475;%%
\lref\fl{V.S.~Fadin and L.N.~Lipatov,
\PL\vyp{B429}{1998}{127}.  %%CITATION = PHLTA,B315,157;%%}
}
\lref\fleta{
V.S.~Fadin et al, \PL\vyp{B359}{1995}{181}; %%CITATION = PHLTA,B359,181;%%
\PL\vyp{B387}{1996}{593}; %%CITATION = PHLTA,B387,593;%%
\NP\vyp{B406}{1993}{259}; %%CITATION = NUPHA,B406,259;%%
\PR\vyp{D50}{1994}{5893}; %%CITATION = PHRVA,D50,5893;%%
\PL\vyp{B389}{1996}{737};  %%CITATION = PHLTA,B389,737;%%
\NP\vyp{B477}{1996}{767};  %%CITATION = NUPHA,B477,767;%%
\PL\vyp{B415}{1997}{97};  %%CITATION = PHLTA,B415,97;%%
\PL\vyp{B422}{1998}{287}.} %%CITATION = PHLTA,B422,287;%%
\lref\cc{G.~Camici and M.~Ciafaloni,
\PL\vyp{B412}{1997}{396}; %%CITATION = PHLTA,B412,396;%%
\PL\vyp{B430}{1998}{349}.} %%CITATION = PHLTA,B430,349;%%
\lref\dd{V.~del~Duca, \PR\vyp{D54}{1996}{989};%%CITATION = PHRVA,D54,989;%%
\PR\vyp{D54}{1996}{4474}\semi %%CITATION = PHRVA,D54,4474;%%
V.~del~Duca and C.R.~Schmidt,
\PR\vyp{D57}{1998}{4069}\semi %%CITATION = HEP-PH 9711309;%%
Z.~Bern, V.~del~Duca and C.R.~Schmidt,
\PL\vyp{B445}{1998}{168}.}%%CITATION = PHLTA,B445,168;%%
\lref\ross{
D.~A.~Ross,
Phys.\ Lett.\ B {\bf 431}, 161 (1998) %%CITATION = HEP-PH 9804332;%%
}
\lref\jar{T.~Jaroszewicz,
\PL\vyp{B116}{1982}{291}.}%%CITATION = PHLTA,B116,291;%%
\lref\ktfac{S.~Catani, F.~Fiorani and G.~Marchesini,
\PL\vyp{B336}{1990}{18}\semi %%CITATION = PHLTA,B336,18;%%
S.~Catani et al.,
\NP\vyp{B361}{1991}{645}.}%%CITATION = NUPHA,B361,645;%%
\lref\summ{R.~D.~Ball and S.~Forte,
\PL\vyp{B351}{1995}{313}\semi  %%CITATION = PHLTA,B351,313;%%
R.K.~Ellis, F.~Hautmann and B.R.~Webber,
\PL\vyp{B348}{1995}{582}.}%%CITATION = PHLTA,B348,582;%%
\lref\afp{R.~D.~Ball and S.~Forte,
\PL\vyp{B405}{1997}{317}.}%%CITATION = PHLTA,B405,317;%%
\lref\DGPTWZ{A.~De~R\'ujula {\it et al.},
\PR\vyp{D10}{1974}{1649}.}%%CITATION = PHRVA,D10,1649;%%
\lref\das{R.~D.~Ball and S.~Forte,
\PL\vyp{B335}{1994}{77}; %%CITATION = PHLTA,B335,77;%%
\vyp{B336}{1994}{77}\semi %%CITATION = PHLTA,B336,77;%%
{\it Acta~Phys.~Polon.~}\vyp{B26}{1995}{2097}.}%%CITATION = HEP-PH 9512208;%%
\lref\kis{
See {\it e.g.}  R.~K.~Ellis, W.~J.~Stirling and B.~R.~Webber,
``QCD and Collider Physics'' (C.U.P., Cambridge 1996).}
\lref\hone{H1 Collab., {\it Eur.\ Phys.\ J.} C {\bf 21} (2001)
33.}%%CITATION = HEP-EX 0112053;%%
\lref\zeus{ZEUS Collab., {\it Eur.\ Phys.\ J.}
 C {\bf 21} (2001) 443.} %%CITATION = HEP-EX 0105090;%%
\lref\mom{R.~D.~Ball and S.~Forte, {\it Phys. Lett.} {\bf
B359}, 362 (1995).}%%CITATION = PHLTA,B359,362;%%
\lref\bfklfits{R.~D.~Ball and S.~Forte,
{\tt hep-ph/9607291}\semi %%CITATION = HEP-PH 9607291;%%
I.~Bojak and M.~Ernst, \PL\vyp{B397}{1997}{296};%%CITATION = PHLTA,B397,296;%%
\NP\vyp{B508}{1997}{731}\semi%%CITATION = NUPHA,B508,731;%%
J.~Bl\"umlein  and A.~Vogt,
\PR\vyp{D58}{1998}{014020}.}%%CITATION = PHRVA,D58,014020;%%
\lref\flph{R.~D.~Ball  and S.~Forte,
{\tt hep-ph/9805315}\semi %%CITATION = HEP-PH 9805315;%%
J. Bl\"umlein et al.,
{\tt hep-ph/9806368}.}%%CITATION = HEP-PH 9806368;%%
\lref\salam{G.~Salam, \JHEP\vyp{9807}{1998}{19}.}%%CITATION = HEP-PH 9806482;%%
\lref\sxap{R.~D.~Ball and S.~Forte,
\PL\vyp{B465}{1999}{271}.}%%CITATION = PHLTA,B465,271;%%
\lref\sxres{G. Altarelli, R.~D. Ball and S. Forte,
\NP{\bf B575}, 313 (2000);  %%CITATION = HEP-PH 9911273;%%
see also {\tt hep-ph/0001157}.%%CITATION = HEP-PH 0001157;%%
}
\lref\bateman{G. Altarelli, R.~D. Ball and S. Forte,
CERN-PH-TH-2005-174, {\tt hep-ph/0001157}.}
\lref\sxphen{G. Altarelli, R.~D.~Ball and S. Forte,
\NP\vyp{B599}{2001}{383};  %%CITATION = HEP-PH 9911273;%%
see also {\tt hep-ph/0104246}.}  %%CITATION = HEP-PH 0104246;%%
\lref\ciaf{M.~Ciafaloni and D.~Colferai,
\PL\vyp{B452}{1999}{372}. %%CITATION = PHLTA,B452,372;%%
}
\lref\ciafresa{
 M.~Ciafaloni, D.~Colferai and G.~P.~Salam,
  %``Renormalization group improved small-x equation,''
  Phys.\ Rev.\ D {\bf 60} (1999) 114036.
  %%CITATION = HEP-PH 9905566;%%
}
\lref\ciafresb{M.~Ciafaloni, D.~Colferai, G.~P.~Salam and A.~M.~Stasto,
  %``Expanding running coupling effects in the hard pomeron,''
  Phys.\ Rev.\ D {\bf 66} (2002) 054014; 
  %%CITATION = HEP-PH 0204282;%%%``Extending QCD perturbation theory to higher energies,''
  Phys.\ Lett.\ B {\bf 576} (2003) 143;
  %%CITATION = HEP-PH 0305254;%%
  %``Renormalisation group improved small-x Green's function,''
  Phys.\ Rev.\ D {\bf 68} (2003) 114003.
  %%CITATION = HEP-PH 0307188;%%
}\lref\heralhcres{G.~Altarelli {\it et al.}, ``Resummation'', in M.~Dittmar {\it et al.},
  %``Parton distributions: Summary report for the HERA - LHC workshop,''
  {\tt hep-ph/0511119.}}
\lref\heralhc{ M.~Dittmar {\it et al.},
  %``Parton distributions: Summary report for the HERA - LHC workshop,''
  {\tt hep-ph/0511119.}
  %%CITATION = HEP-PH 0511119;%%
}
\lref\ciafdip{M.~Ciafaloni, D.~Colferai, G.~P.~Salam and A.~M.~Stasto,
  %``The gluon splitting function at moderately small x,''
  Phys.\ Lett.\ B {\bf 587} (2004) 87; see also
  %%CITATION = HEP-PH 0311325;%%
 G.~P.~Salam,
  %``Asymptotics and preasymptotics at small x,''
{\tt hep-ph/0501097}.
  %%CITATION = HEP-PH 0501097;%%

}
\lref\sxsym{G.~Altarelli, R.~D.~Ball and S.~Forte,
  %``Perturbatively stable resummed small x evolution kernels,''
 {\tt hep-ph/0512237}.
  %%CITATION = HEP-PH 0512237;%%
}
\lref\Liprun{L.N.~Lipatov,
\SPJETP\vyp{63}{1986}{5}.}%%CITATION = SPHJA,63,5;%%
\lref\ColKwie{
J.~C.~Collins and J.~Kwiecinski, \NP\vyp{B316}{1989}{307}.}%%CITATION =
%%NUPHA,B316,307;%%
\lref\CiaMue{
M.~Ciafaloni, M.~Taiuti and A.~H.~Mueller,
%``Diffusion corrections to the hard pomeron,''
{\tt hep-ph/0107009}.
%%CITATION = HEP-PH 0107009;%%
}
\lref\ciafac{
M.~Ciafaloni, D.~Colferai and G.~P.~Salam,
%``On factorization at small x,''
JHEP {\bf 0007}  (2000) 054
%%CITATION = HEP-PH 0007240;%%
}
\lref\ciafrun{G.~Camici and M.~Ciafaloni,
\NP\vyp{B496}{1997}{305}.}%%CITATION = NUPHA,B496,305;%%
\lref\Haak{L.~P.~A.~Haakman, O.~V.~Kancheli and
J.~H.~Koch \NP\vyp{B518}{1998}{275}.} %%CITATION = NUPHA,B518,275;%%
\lref\Bartels{N. Armesto, J. Bartels and M.~A.~Braun,
\PL\vyp{B442}{1998}{459}.} %%CITATION = PHLTA,B442,459;%%
\lref\Thorne{R.~S.~Thorne,
\PL\vyp{B474}{2000}{372}; {\it Phys.\ Rev.} {\bf D64} (2001) 074005.
%%CITATION = PHLTA,B474,372;%%
%%CITATION = HEP-PH 0103210;%%
}
\lref\anders{
J.~R.~Andersen and A.~Sabio Vera,
{\tt arXiv:hep-ph/0305236.}
%%CITATION = HEP-PH 0305236;%%
}
\lref\mf{
S.~Forte and R.~D.~Ball,
%``Theoretical aspects of HERA physics,''
AIP Conf.\ Proc.\  {\bf 602} (2001) 60
{\tt hep-ph/0109235.}
%%CITATION = HEP-PH 0109235;%%
}
\lref\sxrun{
G.~Altarelli, R.~D.~Ball and S.~Forte,
%``Factorization and resummation of small x scaling violations with  running
%%coupling,''
Nucl.\ Phys.\ B {\bf 621} (2002)  359.
%%CITATION = HEP-PH 0109178;%%
}
\lref\ciafqz{
  M.~Ciafaloni,
  %``k(T) factorization versus renormalization group: A Small x consistency
  %argument,''
  Phys.\ Lett.\ B {\bf 356}, 74 (1995)
  %%CITATION = HEP-PH 9507307;%%
}
\lref\ciafrc{M.~Ciafaloni, M.~Taiuti and A.~H.~Mueller,
{\it Nucl.\ Phys.}  {\bf B616} (2001) 349\semi %%CITATION = HEP-PH 0107009;%%
M.~Ciafaloni et al., {\it Phys. Rev.} {\bf D66} (2002)
054014 %%CITATION = HEP-PH 0204282;%%
}
\lref\runph{G.~Altarelli, R.~D.~Ball and S.~Forte,
%``An anomalous dimension for small x evolution,''
Nucl.\ Phys.\ B {\bf 674} (2003) 459;
%%CITATION = HEP-PH 0306156;%%
see also   %``An improved splitting function for small x evolution,''
 {\tt hep-ph/0310016.}
  %%CITATION = HEP-PH 0310016;%%
}
\lref\sxsym{G.~Altarelli, R.~D.~Ball and S.~Forte,
  %``Progress in the small x resummation of the singlet anomalous dimension,''
  Nucl.\ Phys.\ Proc.\ Suppl.\  {\bf 135} (2004) 163
  %%CITATION = HEP-PH 0407153;%%
}
\lref\nnlo{ S.~Moch, J.~A.~M.~Vermaseren and A.~Vogt,
  %``The three-loop splitting functions in QCD: The singlet case,''
  Nucl.\ Phys.\ B {\bf 691} (2004) 129\semi
  %%CITATION = HEP-PH 0404111;%%
 A.~Vogt, S.~Moch and J.~A.~M.~Vermaseren,
  %``The three-loop splitting functions in QCD: The non-singlet case,''
  Nucl.\ Phys.\ B {\bf 688} (2004) 101.
  %%CITATION = HEP-PH 0403192;%%
}
\lref\mrst{See e.g. R.~S.~Thorne, A.~D.~Martin, R.~G.~Roberts and W.~J.~Stirling,
  %``Recent progress in parton distributions and implications for LHC physics,''
  {\tt hep-ph/0507015.}
  %%CITATION = HEP-PH 0507015;%%
}
\newsec{BFKL evolution and GLAP evolution}
\noindent
The all-order summation of large logarithms of the center-of-mass energy in
perturbative QCD is an
intrinsically intricate problem, because it requires the simultaneous
resummation of two large scales: the large energy, and the scale
which renders the strong coupling perturbative. This difficulty is
apparent when trying to fully include the perturbative running of the
coupling in the evolution equation which sums energy logs in
perturbative QCD, the BFKL equation. In fact, doubts have been raised
in the past on the consistency of a BFKL equation with running
coupling. It is also apparent when trying to include a high-energy
(i.e. small $x$) resummation in the standard renormalization-group (or
GLAP) evolution equations of perturbative QCD: indeed, until recently
it was not clear whether (beyond the leading-log
level) a resummation of  energy logs would be possible using
the GLAP equation, without having to resort to the BFKL equation.

The perturbative expansion of the singlet GLAP
splitting functions in powers of $\as(Q^2)$ at fixed $x$ is unstable
at small $x$: indeed,
the recently computed NNLO corrections 
show explicitly that the instability is visible 
already around $x\sim 10^{-3}$. Analogously, the perturbative
expansion 
of the BFKL kernel in powers of $\as(Q^2)$ at fixed $Q^2$ is unstable
at large $Q^2$. In fact, as
 well
known, this instability in practice makes the perturbative expansion
of the BFKL kernel very poorly behaved essentially for all $Q^2$.
Fortunately, it has now been demonstrated 
explicitly~\refs{\sxres\sxsym\bateman\ciafresa\ciafresb{--}\heralhcres}  that
simultaneous resummation of the two large logs is feasible, and cures
both instabilities. This double resummation can be accomplished in a
transparent way by exploiting the
duality~\refs{\afp\sxap\sxres{--}\sxrun} which relates the BFKL and
GLAP equations, whereby one shows that, at the leading-twist level, 
both equations admit the same solutions and thus describe the same
kind of evolution, provided their respective kernels are suitably
matched.

Whereas at the fixed--coupling level this duality is straightforward
to prove, it becomes rather nontrivial at the running
coupling level. In fact, even the
validity of running--coupling factorization at small $x$ to any given
logarithmic order (i.e. the
existence of a universal evolution kernel which sums all pertinent
logs for any given boundary condition) is nontrivial. Running 
coupling factorization and duality to any
logarithmic order at small $x$ were first shown in Ref.~\sxrun. The
approach of that reference, however, besides being rather formal,
was essentially geared to computation of small $x$ corrections to
the GLAP equation, given the BFKL kernel as input. The inverse
problem, namely the computation of large $Q^2$ corrections to the BFKL
kernel given the GLAP kernel, though in principle possible, in
practice requires an all-order computation. This computation is 
required in order to calculate stable
BFKL kernels, and thus essential for any kind of phenomenology based
on the BFKL equation.  Furthermore, in the approach of Ref.~\sxrun\
even the determination of 
next-to-next-to-leading $\ln {1\over x}$ running-coupling corrections 
to the GLAP kernel obtained by fixed--coupling duality from BFKL  is
rather cumbersome.

In this paper, we will develop a rather general 
approach to running-coupling duality, based on the observation that
the standard duality relations between GLAP and BFKL kernels, which
are algebraic equations between kernels in a suitable
double-Mellin transformed space, are promoted to operator equations
when the coupling runs. We will show that these equations can be
solved order-by-order by purely algebraic methods in terms of the
commutators of the relevant operators, without having to actually
determine their spectra, i.e. without having to solve high-order
differential equations.  
This will enable us to determine
running-coupling corrections to duality in either direction (i.e. when
the BFKL kernel is determined from the GLAP one or conversely) and for
either kind of logs. We will check the method
and display its power by reproducing known results in a
compact and efficient way, and then we will use
it to obtain some new all-order results. As a
byproduct, we will obtain a very direct re-derivation of 
factorization at small $x$.

The paper is organized as follows. In Sect.~3 we will define our
notation, recall  some
basic results about duality at fixed and running coupling, and
introduce the general formalism for the representation of running
coupling evolution equations as operator equations. 
Then, in Sect.~4 we will derive within the operator approach
the small $x$ resummation of the GLAP kernel from the BFKL kernel
which was already discussed in Ref.~\sxrun. We will then derive a compact
expression, based on the Baker-Campbell-Hausdorff formula, which
generates the corrections to fixed--coupling duality to any desired
order (and with the running of the coupling given to any desired
order). We will check it by reproducing
the next-to-next-to-leading running coupling corrections to duality
which in Ref.~\mf\ could only be determined using computer algebra.
Finally, in Sect.~5 we will construct
the large $\ln Q^2$ resummation of
the BFKL kernels from the GLAP kernels, which requires the computation
of an infinite set of running-coupling terms: first, we will show that at the
leading--log level there exists an operator ordering such that
running--coupling duality relations reduced to the naive
fixed--coupling ones, then, we will show that
beyond the leading log level naive duality is 
modified by commutator terms, which we will 
determine explicitly up to the
next-to-leading level. This will provide the first determination of a
full collinear resummation of the next-to-leading-order BFKL
kernel. 

\newsec{Duality with  fixed coupling and running coupling}

\subsec{General formalism}
\noindent
We will discuss evolution equations for a parton distribution
$G = xg(x,Q^2)$. 
This, according to the definition of the
factorization scheme, can be thought of as the gluon density, or as
an  eigenvector of a two-by-two evolution matrix in the singlet
sector. The kinematic variables $x$ and $Q^2$ can be thought of as the
standard DIS variables, or more generally the perturbative scale and
a dimensionless scale ratio such that $0\le x\le 1$: for example, for
the production of a final state $X$ in hadronic collisions
with center-of-mass energy $s$, $Q^2\equiv
M^2_X$ (e.g. the Higgs mass), and $x\equiv Q^2/s$.
We will not consider the dependence on other kinematic variables, such
as transverse momentum and rapidity, i.e. we will consider standard
parton distributions, for which ordinary collinear factorization
applies.

In practice, it is convenient to express the parton density as a
function of the logs of the relevant kinematic variables:
\eqn\baskin
{G=G(\xi,t);\qquad \xi\equiv\log{1\over x},\quad
t\equiv \log{Q^2\over \mu^2}. }
We define the Mellin transform with respect to either (or
both) of the kinematic variables:\eqnn\nmom\eqnn\mmom
$$ \eqalignno{G(N,t) &\equiv\int^{\infty}_{0}\! d\xi\, e^{-N\xi}
 G(\xi,t),  &\nmom\cr
 G(\xi,M)&\equiv\int^{\infty}_{-\infty}\! dt\, e^{-Mt}
 G(\xi,t). &\mmom\cr 
}$$
Note that, by slight abuse of notation, we denote with the same
 symbol the parton distributions $G(N,t)$,  $G(\xi,M)$,  and $G(N,M)$,
 although they are of course different functions of the respective
 arguments. In this paper we will mostly deal with the
 double--transformed distribution $G(N,M)$.

The GLAP equation takes the form 
\eqn\tevol{
\frac{d}{dt}G(N,t)=\gamma(\as(t),N) G(N,t), }  where $\as(t)$ is the running
coupling, while the BFKL equation at the {\it fixed coupling} level
takes the form
\eqn\xevol{
\frac{d}{d\xi}G(\xi,M)=\chi(\as,M) G(\xi,M). }
If we expand the GLAP anomalous dimension $\gamma$ and the BFKL
kernel $\chi$ in powers of $\as$ keeping the respective kinematic
variable ($N$ or $M$) fixed \eqnn\gamexp\eqnn\chiexp
$$\eqalignno{\gamma(\as,N)=&\alpha_s \gamma_0(N)~+~\alpha_s^2
\gamma_1(N)~+\dots &\gamexp\cr
\chi(\as,M)=&\alpha_s \chi_0(M)~+~\alpha_s^2 \chi_1(M)~+~\dots 
,&\chiexp}$$ 
the GLAP equation~\tevol\ sums N$^{k+1}$LO $\ln Q^2$ terms if contributions up
to $\gamma_k$ are included, and
the BFKL equation~\xevol\ sums N$^{k+1}$LO $\ln {1\over x}$ terms if
contributions  up
to $\chi_k$ are included in the respective kernels.
 
The {\it fixed coupling} 
duality relation states that if kernels $\chi$ and $\gamma$
related by
\eqnn\dualdef\eqnn\dualinv
$$\eqalignno{ 
N&=\chi(\as,\gamma(\as,N)),&\dualdef\cr
M&=\gamma(\as,\chi(\as,M), &\dualinv\cr
}
$$
then the BFKL and GLAP equation admit the same solution. This relation
is straightfoward to derive~\refs{\sxres,\sxap} from the observation that
the leading twist 
behaviour of $G(N,M)$ is determined by its pole in the $M,N$ plane,
and that eqs.~\dualdef,\dualinv\ express the position of this pole.

\subsec{Running coupling duality} 
\noindent
At the running coupling level, the BFKL kernel becomes an
operator in $M$ space. Indeed, the $(x,Q^2)$ BFKL equation, obtained
by inverse Mellin transform from eq.~\xevol\ has the form
\eqn\qbfkl{\frac{d}{d\xi}G(\xi,Q^2)=\int_{-\infty}^{\infty}
\frac{dk^2}{k^2}  \sum_{n=0}^\infty \as^n K_n(Q^2/k^2)  G(\xi,k^2),}
where
\eqn\bfklker{
\chi_n(M)=\int_{0}^{\infty} \frac{dQ^2}{Q^2}
\left(\frac{Q^2}{k^2}\right)^{-M} 
K_n\left(\smallfrac{Q^2}{k^2}\right).}
Therefore, if in eq.~\bfklker\ we let the coupling run with $Q^2$, 
$\as\to \as(Q^2)$, upon
Mellin transformation we get
\eqn\rcbfkl{
\frac{d}{d\xi}G(\xi,M)= \sum_{n=0}^\infty \ahat^n
\chi_n(M) G(\xi,M),}
where $\ahat$ is the operator obtained by the replacement $t\to
-{\partial\over \partial M}$ in the expression for  
$\as(Q^2)$.  For example, at the leading log level, where
$\as(Q^2)=\as /(1+\beta_0\as t) $ we
get
\eqn\ashdef{
\ash = \frac{\as}{1-\beta_0 \as \smallfrac{\partial}{\partial M}}, } 
where $\as\equiv\as(\mu^2)$. Note that this implies
\eqn\bascomm{[\ahat^{-1},M]=-\beta_0.}

If we
start again with eq.~\qbfkl, but we let $\as\to\as(k^2)$, upon inverse Mellin
transform we get instead           
\eqn\rcbfklk{\frac{d}{d\xi}G(\xi,M)= \sum_{n=0}^\infty 
\chi_n(M)\ahat^n G(\xi,M).}
Of course, in general $\ahat$ and $\chi_n(M)$ do not commute, and
therefore the explicit form of the functions $\chi_n(M)$ will depend
on the argument of the coupling. However, given $\chi_n(M)$
which correspond to one ordering, those which correspond to a
different ordering can be determined by simply computing the relevant
commutators. Of course, this is just the $M$--space form of the trivial
observation that given the expansion of the kernel
$K(\as(Q^2),Q^2/k^2)$ in powers of $\as(Q^2)$ we may determine its
expansion in terms of $\as(k^2)$ by expressing $\as(Q^2)$ in
terms of $\as(k^2)$ through the renormalization--group equation.
Therefore, at the runnning coupling level the BFKL kernel $\chi(\ahat,M)$
is an operator, and its explicit form depends on the operator
ordering. Note that if the evolution equations are written
in double--Mellin $(M,N)$ space, the GLAP anomalous dimension also
becomes the operator $\gamma(\ahat,N)$ obtained by replacing
$\as(t)\to\ahat$ in eq.~\gamexp, but since $\ahat$ commutes with a
function of $N$ the explicit
form of $\gamma(\ahat,N)$ does not depend on operator ordering.

We will assume henceforth for definiteness that $\as$ runs with
$Q^2$. In such case, the running coupling GLAP and BFKL equations in
$(M,N)$ space are
\eqnn\mnglap\eqnn\mnbfkl
$$\eqalignno{  M G(N,M)&= \gamma(\ahat,N)G(N,M)+ G_0(N)&\mnglap\cr
N G(N,M)&= \chi(\ahat,M)G(N,M)+ F_0(M),&\mnglap\cr
}
$$
where  $\gamma(\ahat,N)$ and $\chi(\ahat,M)$ are respectively defined
by the expansions eq.~\gamexp,\chiexp\ with $\as\to\ahat$ (and all
powers of $\ahat$ to the left), and $G_0$ and $F_0$ are boundary
conditions. 
Running coupling duality is the statement that given $\chi$  there
exists a $\gamma$ such that the solutions to
eqs.~\mnglap,\mnbfkl, when transformed back to $(Q^2,x)$ space,
coincide at leading twist, and conversely given $\gamma$  there
exists a $\chi$ with this property. This result is highly
nontrivial in that it relates solutions to high-order differential
equations. 

With the leading-order form of $\ahat$ eq.~\ashdef, we can formally expand the
running--coupling GLAP and BFKL equations in powers of $\beta_0$. This
generates a series of running--coupling corrections to fixed coupling
duality, which we can compute order by order by solving both equations
iteratively as a series in powers of $\beta_0$, and then matching the
solutions. The fact that this matching is possible is 
nontrivial, however, because if we start, say, from BFKL and determine
the dual GLAP kernel, the $Q^2$ dependence of the solutions must match
exactly (up to higher--twist corrections), i.e. to all orders in an
expansion in powers of $t$ up to the given order in $\beta_0$. So, for
instance, at $O(\beta_0^2)$ the solution contains terms up to $O(t^2)$
in an expansion in powers of $t$, 
which must all be matched in terms of a single $O(\beta_0^2)$ correction
to the duality relation between kernels. 

In Refs.~\refs{\sxap,\mf} it was explicitly checked that this matching
is possible up to order $\beta_0^2$, and the corresponding running
duality corrections were explicitly given for the determination of the
GLAP anomalous dimension from the BFKL kernel. The nontrivial
nature of this matching suggested that running duality should in fact
hold to all orders. In Ref.~\sxrun\ it was shown that this is indeed
the case to all orders in $\beta_0$, 
if one uses the leading--order form eq.~\ashdef\ of
the running coupling, and the leading-order expression of $\chi$
eq.~\chiexp, i.e., if the BFKL kernel is linear in $\ahat$. Indeed, in
this case it is possible to solve the running-coupling $(M,N)$--space
BFKL equation exactly as a differential equation. It can then be
proven that in the large $Q^2$ limit only the inhomogeneous solution
to this differential equation survives, 
which is linear in the boundary condition. Furthemore,  its
$Q^2$--dependence can be determined by saddle--point and shown to
exponentiate in terms of an anomalous dimension which is thereby
determined order-by-order in $\beta_0$ by the saddle point expansion,
and seen to reproduce the perturbative result.

Even though it is reasonable to conjecture that these results are quite 
general, so that factorization and duality hold in fact to any
logarithmic order, no general argument is 
available. Furthermore, the saddle-point method of ref.~\sxrun\ is only
applicable if $\chi$ is linear in $\ahat$ and the latter has the
leading--order form eq.~\ashdef. The perturbative brute--force matching
can of course be applied systematically, but it rapidly becomes
computationally very cumbersome: in fact, even the computation of
$O(\beta_0^2)$ corrections to duality in ref.~\mf\ could only be
accomplished by means of computer algebra. Here we will present an
approach which is entirely general and computationally much simpler,
even though of course still amenable to computer implementation.

\newsec{GLAP leading singularities from BFKL kernels}
\noindent
As apparent from eq.~\ashdef, the running of the coupling is a leading
$\ln Q^2$ effect, but it is subleading $\ln{1\over x}$, and in fact
whereas at leading
$\ln Q^2$ terms to all orders in $\as\beta_0$ contribute, to
next$^n$-to-leading $\ln{1\over x}$ only terms up to
$O((\as\beta_0)^n)$ should be included. Hence, the determination of
running coupling corrections to duality is simpler when computing the
leading and subleading $\ln{1\over x}$ corrections to GLAP splitting 
functions from BFKL kernels than when computing leading and subleading 
$\ln Q^2$ corrections to BFKL kernels from GLAP splitting functions: 
in the former case, the number of terms which must be
determined to each logarithmic order is finite, while it is infinite
in the latter case. In this section we consider the simpler
problem, postponing the more difficult case to the next section.

The problem is to start from eq.~\mnbfkl, with the BFKL
kernel $\chi$ computed  to some finite order in the expansion
eq.~\chiexp, and determine whether there exists a function
$\gamma(\as,N)$ such that, if one uses it in the GLAP equation~\mnglap\
the solutions to the two equations coincide at leading twist. In the
fixed coupling case, the solution is  given by eqs.~\dualdef,\dualinv,
which in particular imply that if terms up to order $\as^n$ are
included in the expansion~\chiexp, then $\gamma$ has the form
\eqn\sxgexp
{\gamma(\as,N)=\gamma_s(N/\as)+\as\gamma_{ss}(N/\as)+\dots.} 

In the running coupling case,
the problem was solved perturbatively in Ref.~\sxap. 
Consider
for definiteness the case in which we only include the
leading--order contribution $\chi_0$ to the expansion of the BFKL
kernel $\chi$ eq.~\chiexp, and we let $\as\to\ahat$ eq.~\ashdef. 
The perturbative solution then proceeds in four steps: first,
 $\ahat$ Eq.~\ashdef\ is expanded
as a power series in $\as\beta_0 \dm$ and the running-coupling BFKL
equation~\mnbfkl\ is solved perturbatively, with a result of the form
\eqn\nmsolrun{
G(N,M)=\left[{1 \over 1- {\as\over N} \chi_0} + 
 \as  \frac{\as}{N}
\Big(\frac{\beta_0\chi_0(\ln\chi_0 G_0)'}{(1- {\as\over N}\chi_0)^2}
+\frac{\beta_0\chi_0 \frac{\as}{N}\chi'_0}
{(1- {\as\over N}\chi_0)^3}\Big)+\ldots  
\right]{G_0(M)\over N}.
}
Second,  the inverse $M$-Mellin transform $G(N,t)$ of this solution is
determined up to the desired order. Third, this solution is compared
to the expansion of the solution to the $(N,t)$ space GLAP
equation~\tevol\ to the same order in $\as$ and linear  in
$t$, and the corresponding
$\gamma$ is determined, as the sum of $\gamma$ determined at the
next-lowest order, and a running-coupling correction. Finally, it
is verified that terms corresponding to all higher order powers of $t$
which appear at the given order in $\alpha$ exponentiate, so that the
solution to Eq.~\tevol\ is fully reproduced to the given order.
 
The exponentiation was verified explicitly up to next-to-next-to
leading order in refs.~\refs{\sxrun,\mf}, and the running coupling anomalous dimension was 
found to be $\gamma(\as(t),N)$, where
\eqn\prcc{
\gamma(\as,N)=\gamma_s(N/\as)+\as\beta_0
\Delta\gamma_{ss}(N/\as)+
(\as\beta_0)^2 \Delta\gamma_{sss}(N/\as) +O(\as\beta_0)^3,}
with $\gamma_s(N/\as)$  determined by fixed coupling duality eq.~\dualdef, 
\ie\ as the solution of 
\eqn\gamessdef{N = \as\chi_0(\gamma_s(N/\as)),} 
and
\eqnn\nlrcc\eqnn\nnlrcc
$$\eqalignno{\Delta\gamma_{ss}(N/\as)&=-\chi_0
\frac{\chi_0''}
{2\chi_0'^2}\Bigg\vert_{M=\gamma_s(N/\as)}&\nlrcc\cr
\Delta\gamma_{sss}(N/\as)&=-\chi_0^2{15{\chi_0''}^3-16
\chi_0'\chi_0''\chi_0'''+3 {\chi_0'}^2\chi_0''''\over 24
{\chi_0'}^5}\Bigg\vert_{M=\gamma_s(N/\as)},&\nnlrcc}$$ 
where all derivatives are with respect to the argument of $\chi_0(M)$, 
which is then evaluated at $\gamma_s(N/\as)$. 

We can view these running coupling terms as a modification of 
$\chi$, which gets replaced by
$\chi^{\rm eff}(\ahat,M)=\ahat\chi_0(M)
+\ahat^2 (\chi_1(M)+\Delta\chi_1(M))+\dots$ from
which $\gamma$ eq.~\prcc\ is obtained using naive duality
eq.~\dualdef, with 
\eqn\chioneeff{\Delta\chi_1=\beta_0
\frac{\chi_0''\chi_0}{2\chi_0'},}
and so on. 
Of course, if higher order contributions to $\chi$ eq.~\chiexp\
are included, they also  contribute to $\gamma_s$, $\gamma_{ss}$, etc.,
as dictated by fixed--coupling duality.

This next-to-next-to-leading order  exponentiation was shown to follow
from an all--order argument in ref.~\sxrun, were the next-to-leading
correction was also rederived as a byproduct through a different,
though hardly easier method, based on an asymptotic expansion of the
exact solution of the running coupling BFKL equation~\mnbfkl . 
Here we will show that these results can be obtained in a much simpler
and general algebraic setting without having to actually solve the running
coupling BFKL and GLAP equations, either perturbatively or exactly.

\subsec{A model calculation} 
\noindent
Let us start by considering a toy calculation where the GLAP kernel is
given by its leading singularity, \ie\ 
$\gamma(\ahat,N)=\Ni  \ahat $. In this case, the GLAP
eq.~\mnglap\ reduces to
\eqn\toyglap{\left(M- \Ni \ahat\right) G(N,M)= G_0(N),}
which, multiplying by $N \Mi$ on the left and noting that
$[N,\ahat]=[N,M]=0$  reduces to
\eqn\toyglapb{\left(N-  \Mi\ahat \right) G(N,M)= \Mi N G_0(N).}
Recalling~\refs{\sxres,\sxap} that the large $Q^2$ behaviour of 
the solution is determined by the pole in the $M,N$ plane, this is seen to
reduce to the BFKL-like equation
\eqn\totbfkl{NG(N,M)= \Mi\ahat G(N,M)+ F_0(M),}
with a suitably readjusted boundary condition 
$F_0(M)=M^{-2}\ahat G_0(\Mi\ahat)$. 
Hence, the running dual of $\gamma=\Ni\ahat$ is
\eqn\toycomm{\eqalign{\chi(\ahat,M)&=\Mi\ahat=\ahat \Mi + \beta_0\Mi \ahat^2
\Mi\cr &=\ahat \Mi + \beta_0\ahat^2 M^{-2}+O(\ahat^3),}}
where we have used the commutator
\eqn\invbascomm{[\ahat,M^{-1}]=-\beta_0\ahat M^{-2}\ahat
=-\beta_0\Mi\ahat^2\Mi,}
which may be readily obtained from
eq.~\bascomm.

Now we note that the $O(\ahat)$ term is just the fixed-coupling dual
of $\gamma$, while the $O(\ahat^2)$ term coincides with minus the running
coupling correction eq.~\chioneeff: with $\chi_0=1/M$,
$\frac{\chi_0''\chi_0}{2\chi_0'}=- 1/M^{2}$. This is as it should be, because
Eq.~\nlrcc\ gives
the addition to $\gamma$ due to the running-coupling corrections
which are necessary for it to be the running coupling dual of a given
$\chi$. 
Equation~\chioneeff\ then expresses this quantity as an effective
contribution to $\chi$. Equation~\toycomm\ instead gives the 
$\chi$ which we must start from in order to obtain $\gamma=\as/N$
through running-coupling duality. Clearly, this $\chi$ is obtained by
undoing the effect of running duality, i.e. by starting off with a
$\chi$ which differs from the naive dual by a term which is then
removed by the running coupling correction.

We conclude that in this simple case we have in fact reproduced the
running-coupling correction to duality. We notice furthermore that the
boundary condition did not play any role in the argument: in fact, the
above argument can be recast as the simple observation that
the operator equation
\eqn\toypole{NG(N,M)= \Mi\ahat G(N,M),}
can be rewritten as 
\eqn\revtoypole{MG(N,M)=\Ni\ahat G(N,M) . }
The running-coupling duality corrections are then simply due to the
commutators which are necessary in order to bring the
the BFKL kernel in the form  of eq.~\rcbfkl, which in this simple
example are straightforward to compute.

In the next subsection we shall prove that in fact the determination of
running coupling duality corrections can always be reduced to the 
algebraic manipulation of an operator equation, which we will then 
perform systematically in the subsequent section.

\subsec{Factorization}
\noindent
In order to discuss the form of the duality condition, it is more
convenient to start from the GLAP equation~\mnglap, which can be
formally solved as
\eqn\forsol{G(N,M)=\left( M - \gamma(\ahat,N)\right)^{-1}
G_0(N).}
This equation formally embodies all order (GLAP) factorization, in the 
sense that on the \rhs\ the perturbative evolution factorises from 
the nonperturbative boundary condition $G_0(N)$. We can expand this solution
about any $\bar N$ as 
\eqn\nexp{G(N,M)= \left[M - \gamma(\ahat,\bar{N})
-\gamma'(\ahat,\bar{N})(N-\bar{N})
+O((N-\bar{N})^2)\right]^{-1}(G_0(\bar{N})+O(N-\bar{N})),}
where the prime denotes partial differentiation with respect to $N$.

Now, the leading twist behaviour of $G(N,t)$ at large $t$ is given by the 
pole in the $M$-plane due to the vanishing of the denominator in \forsol:
at the pole 
\eqn\polecg{ MG(N,M)=\gamma(\ahat,N)G(N,M).}
Clearly the position of the pole will depend on $N$. Now 
assume that there exists an operator-valued function $\chi(\ahat,M)$ such 
that at the pole eq.~\polecg\ may be inverted, \ie\
\eqn\polecc{ NG(N,M)=\chi(\ahat,M)G(N,M)}
The operator $\chi(\ahat,M)$ is the inverse of the operator $\gamma(\ahat,N)$,
in the sense that
\eqn\chiopd{MG(N,M)=\gamma(\ahat,\chi(\ahat,M))G(N,M),}
and conversely
\eqn\gamopd{NG(N,M)=\chi(\ahat,\gamma(\ahat,N))G(N,M).}
Of course the existence of the function $\chi(\ahat,M)$ is nontrivial: 
we will show how it may be constructed order by order in the next 
section. 

Given $\chi(\ahat,M)$, we may then choose $\bar N$ in the 
expansion \nexp\ to be the position of the perturbative pole in the $N$-plane
for a given $M$: using \chiopd\ the first two terms in 
the square brackets then cancel by construction, and we are left with
\eqn\expsol{G(N,M)=\left(N-\chi(\ahat,M))\right)^{-1}
(-\gamma'(\ahat,\chi(\ahat,M)))^{-1}G_0(\chi(\ahat,M))+\dots,}
where the dots denote terms which are regular in the neighbourhood of 
the pole. We may then define the function
\eqn\newbc{F_0(M)=(-\gamma'(\ahat,\chi(\ahat,M)))^{-1}G_0(\chi(\ahat,M)),}
which is just some function of $M$, related in a complicated
way to the original function $G_0(M)$, but with manifestly 
no dependence on $N$. Identifying $F_0(M)$ with a BFKL boundary 
condition, eq.~\expsol\ is the solution to the running coupling 
BFKL equation \mnbfkl, with the
kernel $\chi(\ahat,M)$ given by eq.~\chiopd. The solution is again 
factorised at the pole into a perturbative evolution factor and a 
nonperturbative boundary condition. Thus all orders factorization of
the running coupling BFKL equation follows directly from all orders 
factorization of GLAP provided only that the kernel defined through 
eq.~\chiopd\ actually exists.

%Note that the argument works 
%essentially as in the fixed coupling case thanks to the fact that 
%at the pole all the dependence on the non-commuting operators 
%$\ahat$ and $M$ goes through $\chi(\ahat,M)$, which obviously 
%commutes with itself and with $N$.

It is important to understand in which respect eq.~\chiopd\ differs
from the naive fixed coupling duality relation eq.~\dualinv, to which
it bears formal resemblance. The meaning of this equation is that
at the pole, the operators $N$ and $\chi(\ahat,M)$ have the 
same action on the physical solution $G(N,M)$, even though they are 
in general different operators. The anomalous dimension $\gamma(\ahat,N)$ is
then the operator such that if we apply it to a physical
state at the pole, and then use repeatedly eq.~\polecc, we obtain the l.h.s. of
eq.~\chiopd, \ie\ $M$. Just as in the fixed coupling case the 
duality relation \chiopd\ is invertible: it 
can be used to construct $\chi$
given $\gamma$ or $\gamma$ given $\chi$.  

%Henceforth, we will often omit the function $G(N,M)$ in equations 
%such as \polecc,\polecg,\gamopd,\chiopd, writing instead
%\eqnn\chiopds\eqnn\gamopds
%$$\eqalignno{M&=\gamma(\ahat,N)
%=\gamma(\ahat,\chi(\ahat,M))&\chiopds\cr
%N&=\chi(\ahat,M)=\chi(\ahat,\gamma(\ahat,N))&\gamopds}$$ 
%but it should be understood that these equations mean that the operators on
%either side of the equation have the same actions on the physical 
%solutions $G(N,M)$, not that they coincide in general.

To make this more explicit, consider the simple example of sect.~3.1.
Starting with $\gamma=\Ni\ahat$, the pole condition~\polecg\ becomes
$MG=\Ni\ahat G=\ahat\Ni G$.
Then construct $\chi$ by demanding that it satisfies \chiopd:
\eqn\copex{MG = \ahat(\chi(\ahat,M))^{-1}G.}
Note the ordering here: only when the factor of $\Ni$ is on the right, 
and thus able to act on $G$, can we use \polecc.
Expanding $\chi$ at lowest nontrivial order we get 
\eqn\excomm{\eqalign{MG &=\big[\ahat\left( (\ahat\chi_0(M)  +\ahat^2
\chi_1)+\cdots\right)^{-1}\big]G\cr
&= \big[\ahat\left(\chi_0(M) +\ahat \chi_1(M)
+\cdots\right)^{-1}\ahat^{-1}\big]G\cr
&= \big[\left(\chi_0(M) +\ahat \chi_1(M)+\cdots\right)^{-1} 
-\ahat \beta_0 \frac{\chi'_0(M)}{\chi^2_0(M)}\big]G,}}
where in the last line we used the commutator 
\eqn\iicomm{[\ahati,\chi_0^{-1}(M)]=-\beta_0(\chi_0^{-1}(M))'= 
\beta_0\chi_0'/\chi_0^2.}
We immediately get that to order $\ahat$, $\chi_0(M)=1/M$,
and substituting this back in eq.~\excomm\ and expanding 
$\chi_1(M)=\beta_0/M^2$, in agreement with eq.~\toycomm.

Hence, we see that the running coupling dual is determined by solving
either of the two operator equations~\gamopd,\chiopd\ from the
respective pole conditions~\polecg,\polecc. In the next subsection, we
will construct a solution of these equations which, when expanded out,
generates running coupling corrections to any desired order.

\subsec{A Systematic Algebraic approach}
\noindent
The construction of the running coupling dual of a
given BFKL kernel or GLAP anomalous dimension presented in the
previous subsection can be formalized in the following way: 
given an  operator equation of the form
\eqn\opeq{\hat pG=\hat qG}
(where $G$ denotes the physical solution),
and given a function $f(\hat q)$, determine the function $g$ such that using
eq.~\opeq\ on physical solutions one gets
\eqn\fopeq{f(\hat q)G= g(\hat p)G.} 
Note that $f\neq g$ necessarily, since $\hat p$ and $\hat q$ do not commute: 
for example 
$$\hat q^2G = \hat q \hat p G 
= (\hat p \hat q + [\hat q,\hat p])G = 
(\hat p^2 + [\hat q,\hat p])G \neq \hat p^2 G$$
if $[\hat q,\hat p]\neq 0$. So the problem is in general nontrivial.

Equation~\polecc\ corresponds to the particular choice
\eqn\opiddir{\eqalign{\hat p&=\ahati N\cr
\hat q&=\ahati\chi(\ahat,M),
}}
where for future convenience we have multiplied both sides by
$\ahat$. 
Equation~\polecg\ corresponds instead to  the choice
\eqn\opidinv{\eqalign{\hat p&=M\cr
\hat q&=\gamma(\ahat,N).}}
In the latter case, for example, we can choose the function
$f=\tilde\chi$, such that $\tilde \chi(\gamma(\ahat,N))=N$: because
$\ahat$ and $N$ commute this is just the naive dual. Equation~\fopeq\
then determines the running coupling dual $\chi(M)$ as the function
$g(\hat p)$.

A general solution to this problem may be obtained by using the
Baker-Campbell-Hausdorff formula for a pair of noncommuting operators
$A$ and $B$ which to cubic order is~\bch
\eqn\bch
{e^Ae^B=\exp\{A+B+\smallfrac{1}{2}[A,B]
+\smallfrac{1}{12}([A,[A,B]]+[B,[B,A]])+...\}.}
Letting $A=\hat q$ and $B=\hat p -\hat q$ we get
\eqn\bchpq{e^{\hat{q}}e^{\hat{p}-\hat{q}}=
\exp\{\hat{p}-\smallfrac{1}{2}[\hat{p},\hat{q}]
+\smallfrac{1}{6}[\hat{q},[\hat{q},\hat{p}]]
+\smallfrac{1}{12}[\hat{p},[\hat{p},\hat{q}]] 
+\smallfrac{1}{24}[\hat{q},[\hat{q},[\hat{q},\hat{p}]]]
+\smallfrac{1}{24}[\hat{q},[\hat{p},[\hat{p},\hat{q}]]]+....\}.}
Multiplying the right-hand side by the
identity $e^{\hat{p}}e^{-\hat{p}} = 1$ on the left, and using
the Baker-Campbell-Hausdorff formula again with $A=-\hat p$ and $B$
set equal to the exponent of the exponential on the r.h.s., eq.~\bchpq\  becomes
\eqn\bchpqres{\eqalign{e^{\hat{q}}e^{\hat{p}-\hat{q}}&=e^{\hat{p}}
\exp\{-\smallfrac{1}{2}[\hat{p},\hat{q}]
+\smallfrac{1}{6}[\hat{q},[\hat{q},\hat{p}]]
+\smallfrac{1}{3}[\hat{p},[\hat{p},\hat{q}]]
+\smallfrac{1}{24}[\hat{q},[\hat{q},[\hat{q},\hat{p}]]]\cr
&\qquad\qquad\qquad\qquad\qquad\qquad 
+\smallfrac{1}{8}[\hat{q},[\hat{p},[\hat{p},\hat{q}]]]
-\smallfrac{1}{8}[\hat{p},[\hat{p},[\hat{p},\hat{q}]]]+...\}.}}

Now, eq.~\opeq\ implies that
\eqn\expopeq{ e^{\hat p-\hat q}G= G,}
so eq.~\bchpqres\ gives us an expression for the action of $e^{
\hat q}$ on physical solutions $G$. 
We can use this result to construct a general
solution to our problem by noting that any function $f(\hat q)$ can be
constructed by Taylor expansion as
\eqn\taylor{
f(\hat{q})=e^{\hat{q}\frac{d}{d\lambda}}f(\lambda)\big\vert_{\lambda=0}.}
Rescaling the operators $\hat p$ and $\hat q$ in eq.~\bchpqres\ by
$\frac{d}{d\lambda}$, i.e. letting 
$\hat{p}\rightarrow\hat{p}\frac{d}{d\lambda}$,
$\hat{q}\rightarrow\hat{q}\frac{d}{d\lambda}$ we get finally
\eqn\expfopeq{\eqalign{
f(\hat{q})G&= 
e^{\hat{p}\frac{d}{d\lambda}}
\exp\{-\smallfrac{1}{2}[\hat{p},\hat{q}]\smallfrac{d^2}{d\lambda^2}
+\smallfrac{1}{6}[\hat{q},[\hat{q},\hat{p}]]\smallfrac{d^3}{d\lambda^3}
+\smallfrac{1}{3}[\hat{p},[\hat{p},\hat{q}]]\smallfrac{d^3}{d\lambda^3}
+\smallfrac{1}{24}[\hat{q},[\hat{q},[\hat{q},\hat{p}]]]
\smallfrac{d^4}{d\lambda^4}
\cr
&\qquad\qquad\qquad +\smallfrac{1}{8}[\hat{q},[\hat{p},[\hat{p},\hat{q}]]]
\smallfrac{d^4}{d\lambda^4}
-\smallfrac{1}{8}[\hat{p},[\hat{p},[\hat{p},\hat{q}]]]
\smallfrac{d^4}{d\lambda^4}+...\}
f(\lambda)\big\vert_{\lambda=0}G.}}
Expanding out the exponential on the r.h.s. of eq.~\expfopeq\ leads to
an expression in terms of $f(\hat p)$ and repeated commutators of
$\hat p$ and $\hat q$. For example, treating the simple commutator as
first order, the double commutators as second order and so on,
expanding up to
second order we get
\eqn\eefopeq{\eqalign{
f(\hat{q})&G =
\big\{f(\hat{p})-\smallfrac{1}{2}f''(\hat{p})\left[\hat{p},\hat{q}\right]
+\smallfrac{1}{6}f'''(\hat{p})\left[\hat{q},\left[\hat{q},\hat{p}\right]
\right]\cr
&\qquad\qquad\qquad
+\smallfrac{1}{3}f'''(\hat{p})\left[\hat{p},\left[\hat{p},\hat{q}\right]\right]
+\smallfrac{1}{8}f''''(\hat{p}){\left[\hat{p},\hat{q}\right]}^2
+\ldots)\big\}G.}}
This is the main result of this section. In order to show how it can
be used in practice, in the next subsection we will use it
to reproduce the next-to-next-to-leading order result eq.~\nnlrcc.

\subsec{Running coupling corrections to NNLO}
\noindent
In order to compute running-coupling corrections to the duality
relation between $\chi_0$ and the expansion eq.~\sxgexp\ of $\gamma$
in powers of $\as$ at fixed $\as/N$ we use eq.~\eefopeq\ with the
identification eq.~\opiddir\ of $\hat p$ and $\hat q$. Furthermore, we
assume that $\chi$ is given by its leading order expression,
\eqn\chilo{\chi(\ahat,M)=\ahat \chi_0(M),}
so $\hat q=\chi_0(M)$.
The relevant
commutators are then\eqnn\comm\eqnn\dcomma\eqnn\dcommb
$$\eqalignno{&
\left[\hat p,\hat q\right]=\left[\ahati N,\chi_0\right] = 
-N\beta_0\chi_0'(M)\cr
&\left[\hat p, \left[\hat p,\hat q\right]\right]=
\left[\ahati N,\left[\ahati N,\chi_0(M)\right]\right] = 
(N\beta_0)^2\chi_0''(M)\cr
&\left[\hat q, \left[\hat p,\hat q\right]\right]=
\left[\chi_0(M),\left[\ahati N,\chi_0(M)\right]\right] =0,}$$
where $\chi_0$ and its derivatives are all functions of $M$. 

We now choose as function $f(\hat q)$ the function $\gamma_s$,
which is related to $\chi_0$ by naive fixed-coupling duality 
eq.~\gamessdef, so
\eqn\fcdual{M=\gamma_s(\chi_0(M)).}
Substituting in eq.~\eefopeq\ we get 
\eqn\exprcd{\eqalign{
M = \big\{\gamma_s(\ahati N)
-&\smallfrac{1}{2}\gamma_s''(\ahati N)
\left(-N\beta_0\chi_0'(M)\right)+
\smallfrac{1}{3}\gamma_s'''(\ahati N)
\left((N\beta_0)^2\chi_0''(M)\right)\cr
&+\smallfrac{1}{8}\gamma_s''''(\ahati N)
{\left(-N\beta_0\chi_0'(M)\right)}^2
+O({\hat{\alpha}}^3)\big\}G,}}
where the prime denotes differentiation with respect to $M$ of
$\chi_0$ and with respect to $\ahati N$ of $\gamma_s$. 
Note that in this equation the first term on the r.h.s. is
$O(\ahat^0)$, the second term (coming from the simple commutator in
eq.~\eefopeq)  is
$O(N)=O(\ahat)$ in an expansion in powers of $\ahat$ at fixed
$\ahati N$, and the last two terms, coming from the double commutator
and the  commutator square, are $O(N^2)=O(\ahat^2)$.

The right-hand side of eq.~\exprcd\  does not yet give the sought-for expression for $\gamma$, because
of the residual dependence on $M$ through $\chi_0$ and its
derivatives. This dependence can be eliminated iteratively by solving
the equation to leading order in $\ahat$, back-substituting the result
to determine the NLO solution and so on: to LO of course we simply get
\eqn\losol{MG=\big\{\gamma_s(\ahati N)+O(\ahat)\big\}G,}
which replaced in the $O(\ahat^2)$ term gives 
\eqn\nlosol{
M G   = \left\{\gamma_s(\ahati N)
+\half N\beta_0\gamma_s''(\ahati N)\chi_0'
\left[\gamma_s(\ahati N)\right]+O(\ahat^2)\right\}G.}

Beyond $O(\ahat^2)$ however the back-substitution  becomes nontrivial, 
because an operator function $\hat\O (M)$ on a physical solution on
which $M=\gamma_s(\ahati N )$ does not simply
become $\O(\gamma_s(\ahati N ))$, for precisely the same reason why the
functions $f$ and $g$ in eq.~\fopeq\ are different. Specifically, to
$O(\ahat^2)$ the back-substitution of the leading-order result into
the $O(\ahat)$ term in eq.~\exprcd\ requires the evaluation of
\eqn\backs{
\chi_0'(M)G =\left\{\chi_0'\left[\gamma_s(\ahati N)\right]+\half
N\beta_0\gamma_s'(\ahati N)
\chi_0'''\left[\gamma_s(\ahati N)\right]+O(\ahat^2)\right\}G,}
where the result follows using eq.~\expfopeq\ again, but with $\hat p$
and $\hat q$ given by eq.~\opidinv, and $f=\chi_0'$. Note that it
is sufficient to evaluate eq.~\backs\ to $O(\ahat^2)$ because the term
on the left-hand side
appears in a contribution to  eq.~\exprcd\ which is already $O(\ahat)$.
Using this result, the back-substitution of the next-to-leading order
result eq.~\nlosol\ into the $O(\ahat)$ term in eq.~\exprcd\ has the
form
\eqn\bsfin{\eqalign{\half N\beta_0\gamma_s''\chi_0'(M)G&=\big\{
\half N\beta_0\gamma_s''[\chi_0'(\gamma_s)+
\half N\beta_0\gamma_s'\chi'''_0(\gamma_s)\cr
&\qquad\qquad +\chi''_0(\gamma_s)\half
N\beta_0\gamma_s''\chi_0'(\gamma_s)]+O(\ahat^3)\big\}G,}} 
where the second term in square brackets comes from eq.~\backs, and
the third from the expansion of the argument of $\chi'$
about $\gamma_s$. All terms on the right hand side of this equation 
now depend on $\ahati N$ through $\gamma_s$ and its derivatives.

Collecting everything, the final result for $\gamma$ is given by the
r.h.s. of eq.~\exprcd, with the $O(\ahat)$ term evaluated using
the next-to-leading back-substitution eq.~\bsfin, 
and the two $O(\ahat^2)$ term evaluated with the
leading back-substitution eq.~\losol, to give 
\eqn\nnlosol{\eqalign{
M G &= \big\{\gamma_s
+N\beta_0\half\gamma_s''\chi_0'(\gamma_s)
+(N\beta_0)^2(\quarter (\gamma_s'')^2\chi_0'(\gamma_s)\chi_0''(\gamma_s)
+\quarter\gamma_s'\gamma_s''\chi_0'''(\gamma_s)\cr
&\qquad\qquad \qquad\qquad 
+\smallfrac{1}{3}\gamma_s'''\chi_0''(\gamma_s)
+\smallfrac{1}{8}\gamma_s''''(\chi_0'(\gamma_s))^2)+O(\ahat^3)
\big\}G.}}
In this way, all terms in eq.~\nnlosol\ depend on $\ahati N$ through 
$\gamma_s$ and its derivatives, while the dependence on
$M$ has been eliminated. The result can be further rewritten entirely
in terms of $\chi_0$ and its derivatives, or in terms of $\gamma_s$
and its derivatives, by exploiting the fact that $\gamma_s$ and
$\chi_0$ and their derivatives are related by the fixed coupling 
duality relation eq.~\fcdual\ and the following relations which may be 
obtained differentiating it:
\eqn\dualders{\eqalign{\gamma_s'&= 1/\chi_0',\cr
\gamma_s''&=-\chi_0''/{\chi_0'}^3\cr
\gamma_s'''&=(3 {\chi_0''}^2- \chi_0'\chi_0''')/{\chi_0'}^5\cr
\gamma_s''''&=-(15{\chi_0''}^3-10
\chi_0'\chi_0'' \chi_0'''+{\chi_0'}^2\chi_0''''
\chi_0'')/{\chi_0'}^7,}}
where $\chi_0$ and its derivatives are evaluated as a function of
$\gamma_s(\ahati N)$ (so $\chi_0= \ahati N$).
With these substitutions, eq.~\nnlosol\ takes the form
\eqn\finrcd
{MG = 
\big\{\gamma_s(\ahati N)
+\hat{\alpha}\beta_0\Delta\gamma_{ss}(\ahati N)
+(\hat{\alpha}\beta_0)^2\Delta\gamma_{sss}(\ahati N)+O(\ahat^3)
\big\}G,
}
with $\Delta\gamma_{ss}$ and $\Delta\gamma_{sss}$ given by eq.~\nlrcc\
and eq.~\nnlrcc\ respectively, thus reproducing the result of
ref.~\refs{\sxap,\mf}. 
It is easy to extend this calculation to higher orders, and also to
include the contributions due to higher-order terms in the $\beta$
function, which will simply modify the form  of the basic commutator
eq.~\bascomm. A systematic investigation of such terms is
presented elsewhere~\refs{\fal,\marfal}.~\foot{When going to yet higher
orders two further subtleties appear. First, eq.~\eefopeq\ cannot be
used directly, but rather one must use the modified form of it which 
has the function $f(\hat p)$ all the way to the right in all terms. This
is due to the fact that it is only when acting on the physical solutions
on the right that one can let $\hat p=\hat q$ 
iteratively. The second is that when expanding the argument of
$\chi_0$ and its derivative about $\gamma_s$ it must be taken into
account that $\gamma_s$ is shifted by an amount that does not commute
with $\gamma_s$  itself, so the Baker-Campbell-Hausdorff formula for
the function of the sum of noncommuting operators must be used.}

\newsec{BFKL leading singularities from GLAP anomalous dimensions}
\noindent
Let us now turn to the inverse problem to that which we discussed in
the previous section: namely, the construction of a BFKL kernel which is
dual order by order to a given GLAP anomalous dimension
eq.~\gamexp. Obviously, at
the fixed coupling level, duality maps the
expansion eq~\gamexp\ of $\gamma$ in powers of $\as$ at fixed $N$ 
onto the
expansion of $\chi$ in powers of $\as$ at fixed $\as/M$:
\eqn\chisexp{\chi(\as,M)=\chi_s(M/\as)+\as\chi_{ss}(M/\as)+\dots,}
so that the dual of $\gamma$ when terms up to $\gamma_n$ are included in
eq.~\gamexp\ is given by $\chi$ with terms up to $\chi_{s^n}$ included
in eq.~\chisexp. Now, at the running coupling level, the very definition of
$\chi_s$ requires an operator ordering, since $\ahat$ and $\Mi$ do 
not commute. Moreover, eq.~\invbascomm\ implies that 
different operator orderings within $\chi_s$ differ by
terms which are of the same order as $\chi_s$ itself (and similarly to
subsequent orders). Hence, as already mentioned in section~2, the
determination of the dual $\chi_s$ involves the computation of 
an infinite set of running coupling corrections to duality.
Running duality contributions to $\chi_s$ may be computed to  
any desired order by using the method of the previous section, but
with the identification eq.~\opidinv\ of $\hat p$, $\hat q$. This
may be  useful for the determination~\refs{\marfal,\mar} of 
an approximate form of $\chi_n$
beyond the known next-to-leading order $\chi_1$. 
However, in this section we will show that,
by choosing a suitable operator ordering, it is in fact possible to
determine runnig coupling corrections to all orders in $\as$, and we
will construct explicitly the running coupling dual to $\gamma$
eq.~\gamexp\ up to next-to-leading order.

\subsec{Operator ordering and leading order}
\noindent
In order to introduce our proof,
consider first a simplified case. Namely, assume that we
start with the model leading--order GLAP anomalous dimension
\eqn\toygam
{\gtoy= \as(t) \left(\Ni-1\right).
}
This anomalous dimension has the same small--$N$ behaviour as the
leading--order GLAP anomalous dimension (up to an overall numerical
factor, which is irrelevant here), and it respects momentum
conservation in that it vanishes at $N=1$. In fact, the simple form
eq.~\toygam\ turns out to be a surprisingly good approximation to the
full leading--order GLAP anomalous dimension, even though this is
immaterial for our purposes.   We wish to determine the $\chi_s$
kernel which is dual to it. Because the anomalous dimension is very
simple, we can solve all the relevant evolution equations exactly. 

At the fixed coupling
level, the dual eq.~\dualdef\ of $\gtoy$ is given by
\eqn\toychi{
\ctoy(M/\as)={\as\over\as+M}=\frac{1}{1+M/\as}.} 
We proceed by trial and error: we guess a
particular ordering in $\chi$, we determine the corresponding
$\gamma$ anomalous dimension explicitly, and we compare the result to
eq.~\toygam. Specifically, we make the ansatz
\eqn\ctoyord{\ctoy(\ahati M)\equiv{1\over 1+ \ahati M}={1\over 1+ M \ahati
-\beta_0},}
where in the last step we used the commutator eq.~\bascomm.
This can be thought of as the result of writing the fixed--coupling
dual \toychi\ as a function of $\as^{-1}M$, and then letting $\as\to\ahat$
with this particular ordering.
Substituting in the running-coupling BFKL equation~\mnbfkl\ we get
\eqn\ctoyeq{(1-\beta_0 +M\ahati) N G(N,M)=\tilde G_0(M)+G(N,M),}
where $\tilde G_0(M)=(1-\beta_0 +M\ahati) G(N,M)$ is a modified
boundary condition. We now assume that $\ahat$ takes the
leading--order form eq.~\ashdef. As proven in Ref.~\sxrun, the large
$Q^2$ behaviour of the solution to eq.~\ctoyeq\ is the same as that of
the solution of the associated homogenous equation, multiplied by a
$Q^2$--independent boundary condition. Hence, we determine it
setting $\tilde G=0$ in
Eq.~\ctoyeq, which then has solution
\eqn\ctoysol{G(N,M)=M^{\beta_0^{-1}(1-\Ni)-1}e^{M/\beta_0\as}.}

In order to determine the large $Q^2$ behaviour explicitly, we must
invert the $M$--Mellin transform:
\eqn\ctoysolq{\eqalign{G(N,t)&=\int  \! {dM\over 2\pi i}
M^{\beta_0^{-1}(1-\Ni)-1}e^{M/\beta_0\as} e^{Mt}\cr
&=(\beta_0 \as(t))^{\beta_0^{-1}(1-\Ni)}
[\Gamma\left(\beta_0^{-1}(1-\Ni)-1\right)]^{-1} \cr}.}
The anomalous dimension is then
\eqn\toyadcomp{\gamma(N,\as(t))\equiv\frac{\partial}{\partial t}
\ln G(N,t)=\as(t)\left(\Ni-1\right),}
which coincides with eq.~\toygam. Hence, we find the surprizing result
that for the particular ordering \ctoyord\ the running--coupling dual 
coincides exactly with the naive dual.
 
Let us now prove  that this result holds in general at the
leading $\ln Q^2$ level, provided the operator ordering of
eq.~\ctoyord\ is adopted.
As discussed in sect.~3.2, the dual BFKL kernel is constructed by
starting with the pole condition eq.~\polecg, and determining the
operator $\chi(\ahat,M)$ which satisfies eq.~\gamopd.
Now, at leading order the pole condition is just
\eqn\lopolecg{\ahati M G(N,M)= \gamma_0(N) G(N,M).}
Therefore, given the ordinary inverse function $\chi_s$
\eqn\fcdinv{\chi_s(\gamma_0(N))=N}
then  the operator--valued function $\chi_s(\ahati M)$, when acting on
the solution $G$, using repeatedly eq.~\lopolecg\ is seen to satisfy
\eqn\fcdclo{NG(N,M)=\chi_s(\ahati M)G(N,M). }
Namely, each time $\ahati M$ acts on the physical state $G(N,M)$ it gives
$\gamma(N)$, so acting repeatedly gives powers of the
commuting operator $\gamma(N)$. Therefore if $\chi_s$ is chosen to
satisfy the inverse function condition eq.~\fcdinv\ the desired dual
BFKL equation eq.~\fcdclo\ immediately follows, which is what we set
out to prove. Note that $(\ahati
M)^{-1}=\ahat \Mi$, so we can equivalently take $\chi_s$ to be a function 
of $\ahat\Mi$.

\subsec{Next-to-leading corrections}
\noindent
The result of the previous subsection can be derived equivalently by
letting
\eqn\opidinvmod{\eqalign{
\hat p&=\ahati M\cr
\hat q&=\ahati \gamma(\ahat,N),}}
so that when $\gamma(\ahat,N)=\ahat \gamma_0(N)$, then $\hat p$ 
and $\hat q$ commute, and eq.~\eefopeq\ implies 
that $f(\hat p) G= f (\hat q)G$ 
so taking $f\equiv\chi_s$ defined by
naive duality~\fcdinv\ we get
\eqn\locao{\chi_s(\hat p)G(N,M)=\chi_s(\gamma_0(N))G(N,M),}
whence the result.
However, at next-to-leading order $\hat q=\gamma_0(N)+\ahat
\gamma_1(N)$, and $[\hat p,\hat q]= -\beta_0\ahat\gamma_1(N)$, which is
of the same order as $\gamma_1(N)$. Likewise, all higher order
commutators
$[\hat p,[\hat p,\dots,[\hat p,\hat q]]]$ are of the same order
$\ahat$. Therefore, the iterative commutator approach of the 
previous section does not help and we must proceed to all orders.

This can be done by noting that the desired next-to-leading order
contribution $\chi_{ss}(\ahati M)$ to the BFKL kernel must satisfy [recall
eq.~\chiopd]
\eqn\nloao{\left[\gamma_0(\chi_s(\hp)+\ahat
\chi_{ss}(\hp))+\ahat\gamma_1(\chi_s(\hp))\right]G(N,M)=\hat p G(N,M),}
where $\hp$ is given by eq.~\opidinvmod.
At leading order, this just gives 
\eqn\loao{\gamma_0(\chi_s(\hp))G(N,M)=\hat p G(N,M),}
which is equivalent to eq.~\locao.
In order to determine  the next-to-leading order 
$\chi_{ss}$ it is necessary to expand out the
argument of $\gamma_0$ in eq.~\nloao, which is nontrivial because
$[\ahat,\chi_{ss}(\hp)]\neq 0$. It turns out, however, that 
this commutator can be determined explicitly, and this is sufficient 
to solve eq.~\nloao\ by expansion.

To this end, write $\chi_{ss}$ as a power series:
\eqn\chissexp{\chi_{ss}(\hp)=\sum_i k_i \hp^{-i}.} 
In practice we will find that only positive powers of $i$ contribute
to the sum (since $\chi_{ss}$ vanishes when $\ahat\to0$),
but this is not necessary for our argument. 
Hence, we wish to determine the commutator  $[\hp^{-n},\ahat]$. 
To this purpose, first we note that the commutator eq.~\invbascomm\
implies
\eqn\commapi{[\ahat,\hp^{-1}]=-\beta_0\hp^{-1}\ahat\hp^{-1},}
which can be rewritten as
\eqn\commapibis{\hp^{-1}\ahat
=\ahat\hp^{-1}\left(1-\beta_0\hp^{-1}\right)^{-1}=\ahat(\hp-\beta_0)^{-1}.}
Iterating this equation, it immediately follows that
\eqn\commapin{\hp^{-n}\ahat=\ahat(\hp-\beta_0)^{-n}.}
Similarly, 
\eqn\commap{[\ahat,\hp]=\beta_0\ahat}
implies
\eqn\commapn{\hp^n \ahat=\ahat\hp^n\left(1-\beta_0\hp^{-1}\right)^{n}
=\ahat(\hp-\beta_0)^n,}
or, equivalently, eq.~\commapin\ holds with both positive and negative
$n$. Using this in the expansion eq.~\chissexp\ of $\chi_{ss}$ implies
that
\eqn\chicommapi{
\chi(\hp) \ahat=\ahat\chi\left(\hp-\beta_0\right).}
Of course, this result holds for any function that may be expanded as
a series of powers of $\hp$.

It is now easy to determine the function on the left-hand side of
eq.~\nloao, i.e. $\gamma_0(\chi_s(\hp)+\ahat
\chi_{ss}(\hp))$, by expanding
\eqn\gamoexp{\gamma_0(N)=\sum_j g_j N^j,}
where in practice of course the sum runs over $-1\le j\le\infty$, though 
again we will not actually use this in our argument. 
We must therefore evaluate
\eqn\chipow{\left(\chi_s+\ahat \chi_{ss}\right)^n= \chi^n_s+\left(
\chi^{n-1}_s\ahat \chi_{ss}+\chi^{n-2}_s\ahat
\chi_{ss}\chi_s+\dots+\ahat \chi_{ss}\chi^{n-1}_s\right)+O(\ahat^2) ,} 
where $\chi_s$ and $\chi_{ss}$ are both functions of $\hp$. 
Using eq.~\chicommapi\ we get
\eqn\chipowex{\left(\chi_s+\ahat \chi_{ss}\right)^n= \chi^n_s(\hp)+
\ahat\frac{\chi_s^n(\hp-\beta_0)-\chi_s^n(\hp)}
{\chi_s(\hp-\beta_0)-\chi_s(\hp)}\chi_{ss}(\hp)+O(\ahat^2),}
where we have made use of the fact that
$[\chi_s(\hp-\beta_0), \chi_s(\hp)]=0$
and we have noticed that for a pair of commuting operators $A$ and $B$,
$$A^{n-1}+A^{n-2}B+\dots+B^{n-1}=\frac{A^n-B^n}{A-B}.$$
Furthermore, it is easy to see that eq.~\chipowex\ also holds for
negative $n$, by expanding
\eqn\gamoexpn{\eqalign{
\left(\chi_s+\ahat
\chi_{ss}\right)^{-n}&=\left(\chi_s^{-1}+\chi_s^{-1}\ahat
\chi_{ss}\chi_s^{-1}\right)^{n} +O(\ahat^2)\cr
&=\chi_s^{-n}+\left(\chi_s^{-n}\ahat\chi_{ss}\chi_s^{-1}+\chi_s^{-(n-1)}
\ahat\chi_{ss}\chi_s^{-2}+\dots+\chi_s^{-1}\ahat\chi_{ss} \chi_{s}^{-n}\right)
+O(\ahat^2),}}
and then using again eq.~\chicommapi\ but with $\chi_s\to\chi^{-1}_s$.

Substituting eq.~\chipowex\ in the expansion eq.~\gamoexp\ of $\gamma$
gives
\eqn\expgammaf{\gamma_0(\chi_s(\hp)+\ahat
\chi_{ss}(\hp))=\gamma_0(\chi_s(\hp))+\ahat\chi_{ss}(\hp)
\frac{\gamma_0(\chi_s(\hp-\beta_0))
-\gamma_0(\chi_s(\hp))}
{\chi_s(\hp-\beta_0)-\chi_s(\hp)}}
Using this  in eq.~\nloao\ with the leading--order eq.~\loao\ we
get the remarkably simple result
\eqn\finchiss{\chi_{ss}(\hp)=\beta_0^{-1}\gamma_1(\chi_s(\hp))
\left(
\chi_s(\hp-\beta_0)-\chi_s(\hp)\right).}
This is the running coupling generalization 
of the well-known fixed coupling result 
$\chi_{ss}(M/\as) 
= -\frac{\gamma_1(\chi_s(M/\as))}{\gamma_0'(\chi_s(M/\as))}$, to which 
it reduces in the limit $\beta_0\to 0$ 
(since $\chi_s'=1/\gamma_0'(\chi_s)$).

The result can be generalized to higher orders and systematized by
performing the expansion of the higher--order generalizations of operator 
duality condition eq.~\nloao\
by means of the Baker-Campbell-Hausdorff formula for the Taylor
expansion of a function of the sum of two non-commuting operators. 

\newsec{Conclusions}
\noindent 
In this paper we have presented an explicit constructive proof of the
fact that the leading--twist duality between the BFKL and GLAP
equations holds to all orders, with the running of the coupling
included to any desired order. Besides its conceptual interest, this
result allows one to exploit fully the information contained in
fixed--order determinations of the BFKL and GLAP kernels by using each
to resum the unresummed singularities in the other. 

Because of the instability of the fixed order expansion of the BFKL 
kernel in powers of $\as$ due to unresummed collinear singularities, 
our running duality is especially useful in constructing a resummed 
form of this kernel, using the collinear resummation implicit in 
the fixed order expansion of the GLAP kernel. The running duality can 
then be used again to derive from the resummed kernel a corresponding 
anomalous dimension in which small $x$ singularities are resummed. 
Indeed, several of the results contained in this paper were used already 
in the explicit construction of a fully stable resummed perturbative 
expansion of small $x$ anomalous dimensions~\refs{\sxsym,\bateman,\heralhcres}. 
Further investigations of the application of the formalism presented here 
to this and related problems will be presented elsewhere~\marfal.

\footatend\vfill\supereject\immediate\closeout\rfile\writestoppt
\baselineskip=14pt\centerline{{\bf References}}\bigskip{\frenchspacing%
\parindent=20pt\escapechar=` \input refs.tmp\vfill\eject}\nonfrenchspacing
\vfill\eject
\bye